\documentclass[12pt]{iopart}

\usepackage[style=phys]{biblatex}
\usepackage{graphicx}
\usepackage{dcolumn}
\usepackage{bm}
\usepackage[utf8]{inputenc}
\usepackage[T1]{fontenc}
\usepackage{mathptmx}
\usepackage{siunitx}
\usepackage{xcolor}
\usepackage{amssymb} 
\usepackage{lscape} 
\usepackage{comment}
\usepackage[title]{appendix}
\usepackage{float}

\begin{document}
\title[Operation above density limit in high performance DIII-D NT discharges]{Operation above the Greenwald density limit in high performance DIII-D negative triangularity discharges}

\author{O Sauter$^1$, R. Hong$^2$, A. Marinoni$^{3}$, F. Scotti$^4$, P. H. Diamond$^3$, C. Paz-Soldan$^5$, D. Shiraki$^6$, K. E. Thome$^7$, M. A. Van Zeeland$^7$, H.Q. Wang$^7$, Z. Yan$^8$ and the negD-DIII-D Team$^x$}
\address{$^1$ Ecole Polytechnique Fédérale de Lausanne (EPFL), Swiss Plasma Center (SPC), CH-1015 Lausanne, Switzerland}
\address{$^2$ University of California, Los Angeles, CA, United States of America}
\address{$^3$ University of California, San Diego, CA, United States of America}
\address{$^4$ Lawrence Livermore National Laboratory, Livermore, CA, United States of America}
\address{$^5$ Columbia University, New York, United States of America}
\address{$^6$ Oak Ridge National Laboratory, Oak Ridge, TN, United States of America}
\address{$^7$ General Atomics, San Diego, CA, United States of America}
\address{$^8$ University of Wisconsin-Madison, Madison, WI, United States of America}
\address{$^x$ see K. Thome et al, Plasma Phys. Control. Fusion {\bf 66} (2024) 105018}
\ead{Olivier.Sauter@epfl.ch}

\vspace{10pt}
\begin{indented}
\item[]
\end{indented}

\begin{abstract}
  The density limit in strongly-shaped negative triangularity (NT) discharges is studied experimentally in the DIII-D tokamak. Record-high Greenwald fractions $f_G$ are obtained, using gas puff injection only, with values up to near 2, where $f_G$ is defined as the ratio of the line-averaged density over $n_G=I_p/(\pi\,a^2)$, with $I_p$[MA] the plasma current and $a$[m] the plasma minor radius. A clear higher operational limit with higher auxiliary power is also demonstrated, with the ohmic density limit about two times lower than with additional neutral beam injection heating. The evolution of the electron density, temperature and pressure profiles are analyzed as well. The core density can be up to twice the Greenwald density and keeps increasing, while the value at the separatrix remains essentially constant and slightly below $n_G$. The edge temperature gradient collapses to near zero and NT plasmas are shown to be resilient to such profiles in terms of disruptivity. We also present the time evolution of the inverse electron pressure scale length with the value at the last closed flux surface (LCFS) decreasing below the value at the normalized radius 0.9 near the density limit, demonstrating the clear drop of confinement starting from the edge. This inverse scale length ``collapse'' at the LCFS also defines well the characteristic behavior of the kinetic profiles approaching a density limit.
\end{abstract}

\section{Introduction} 
Dedicated experiments aimed at testing the density limit in strongly-shaped negative triangularity (NT) plasmas and its dependence on auxiliary power input have been performed during the special campaign on the DIII-D tokamak in January 2023. The campaign was relatively short and covered many different reactor relevant topics \cite{Thome2024, Paz-Soldan2024}; therefore only 13 shots were dedicated to the topic of this short paper. These focused on two aspects, the dependence on plasma current, comparing with the Greenwald density limit \cite{Greenwald}, and the dependence of the density limit on input power, as recently proposed as a key physics driver for increasing the density limit \cite{Giacomin2022}. Despite the short campaign, we present unique results that should motivate further dedicated experimental and theoretical studies worldwide. Note that we consider the L-mode density limit, since NT plasmas with such strongly negative top/bottom (-0.35/-0.6) triangularities cannot go into H-mode \cite{NelsonPRL, MarinoniReview}. The power dependence study was inspired by the work of Ref. \cite{Giacomin2022} relating the density limit to a change in the turbulence properties at the edge of the plasma and in the scrape-off layer (SOL).
At high collisionality, turbulence enhances significantly leading to a collapse of the sustained edge temperature gradient. The latter can be kept to a finite value with sufficient heat flux. This is actually used in density limit avoidance scheme \cite{Maraschek, Vu2021}, in X-point radiated scenarios \cite{Stroth} and can be viewed as a global power balance condition between the power crossing the separatrix, ${\rm P_{sep}}$, and the power reaching the divertor \cite{Manz} at the macroscopic level. A recent data-driven cross-machine analysis of positive triangularity (PT) L-mode plasmas highlights the role of the edge collisionality and pressure \cite{Maris_2025}.
Radial transport also depends on other local properties like the local shearing rate \cite{Diamond} and of course on the radiated power \cite{Gates, Zanca} which itself modifies the effective heat flux sustaining the kinetic profile gradients \cite{Giacomin2022, Giacomin2022a}. From these latter considerations, NT plasmas are expected to have a slightly higher density limit thanks to the natural improved confinement, as compared to PT L-mode, allowing access to sustained higher gradients at lower heat fluxes. The main goal of this work was to test this experimentally.

We present the experimental set-up and the main global results at $I_p=0.6$MA in Sec. \ref{sec:scalars}, the kinetic profiles evolution in Sec. \ref{sec:profiles}, the results at higher currents in Sec. \ref{sec:otherIps} and the conclusions in Sec. \ref{sec:conclusion}.

\section{Experimental set-up and main results}
\label{sec:scalars}
The NT operating density range was tested through density ramps at otherwise relatively constant plasma current $I_p$, input power (neutral beam injection (NBI)) $P_{nbi}$ and shape. Only one shape has been used, due to lack of experimental shape development time, the ``campaign shape'' \cite{Thome2024}, a lower single-null diverted shape with top/bottom triangularity near -0.35/-0.6. The aim was twofold, to test the relation with the Greenwald density limit ($n_G=I_p[MA]/\pi/a^2$), since this remains the main experimental scaling in positive triangularity (PT) shapes \cite{Greenwald}, and to probe the effect of additional heating power. We performed 13 similar dedicated discharges with $I_p=0.6 (7), 0.8 (2)\;{\rm and}\;1{\rm MA}\;(4)$, $B_0=+2T$ (Rev $B_0$, ion $B\times \nabla B$ drift away from the X-point) and $P_{nbi}$ between 0 and 12MW. Other discharges, in particular those for detachment studies, reached high densities and complement the present dataset \cite{Scotti2024, Scotti2025}, adding cases in Rev $B_0$ as well. In some cases, the density limit was much higher than anticipated and we had to repeat with a higher maximum density target. In TCV, increasing the core density was sometimes found difficult, instead a density accumulation near the divertor leg was forming \cite{Coda_2022}. In our cases, using a top valve in all cases (hence away from the divertor region), we did not observe any difficulties in increasing the core line-averaged density. As seen from the number of shots mentioned above, we have focused on the case $I_p=0.6$MA, since the other extreme, $I_p=1$MA, was prone to early MHD unrelated to the density limit phenomena (as discussed below) and most probably due to error field effects. The short campaign did not allow for error field corrections to be determined nor optimized.

\begin{figure*}
    \centering
    \includegraphics[scale=0.8]{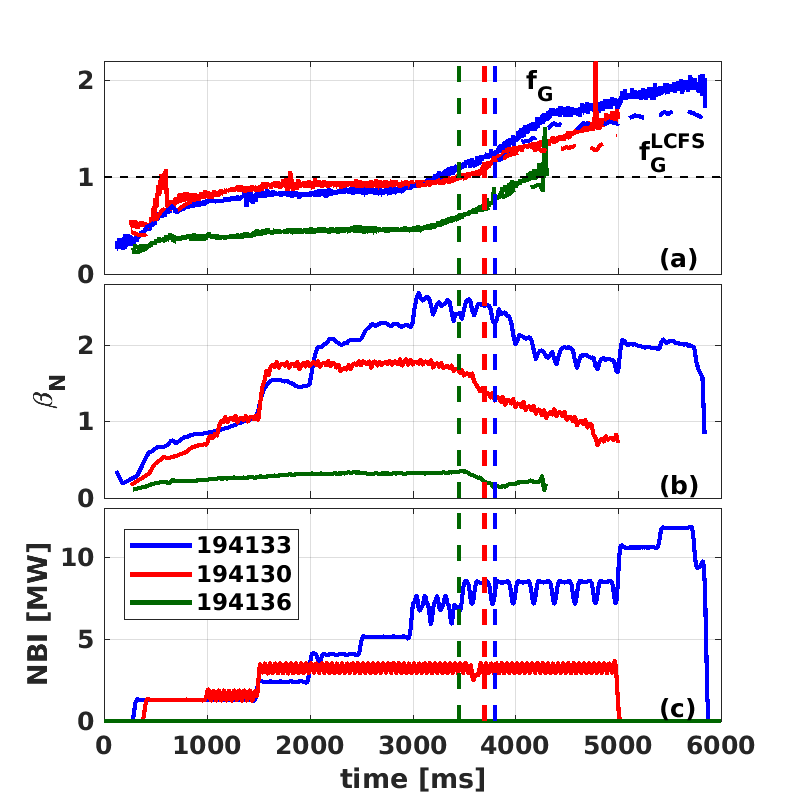}
    \caption{Time evolution of (a) the Greenwald fraction $f_G$ inferred from interferometer (solid lines) and the Greenwald fraction $f_G^{LCFS}$ using only the density profile inside the LCFS from Thomson scattering measurements (dashed lines), (b) $\beta_N$ and (c) NBI power in discharges with similar density ramps at otherwise constant shape and $I_p=0.6$MA, $B_0=+2$T. The horizontal black dashed line highlights $f_G=1$. The approximate times of detachment are marked with vertical dashed lines for 194130  (blue, 3450ms), 194130 (red, 3700ms) and 194133 (blue, 3800ms).}
    \label{fig1}
\end{figure*}

\begin{figure*}
    \centering
    \includegraphics[scale=0.35]{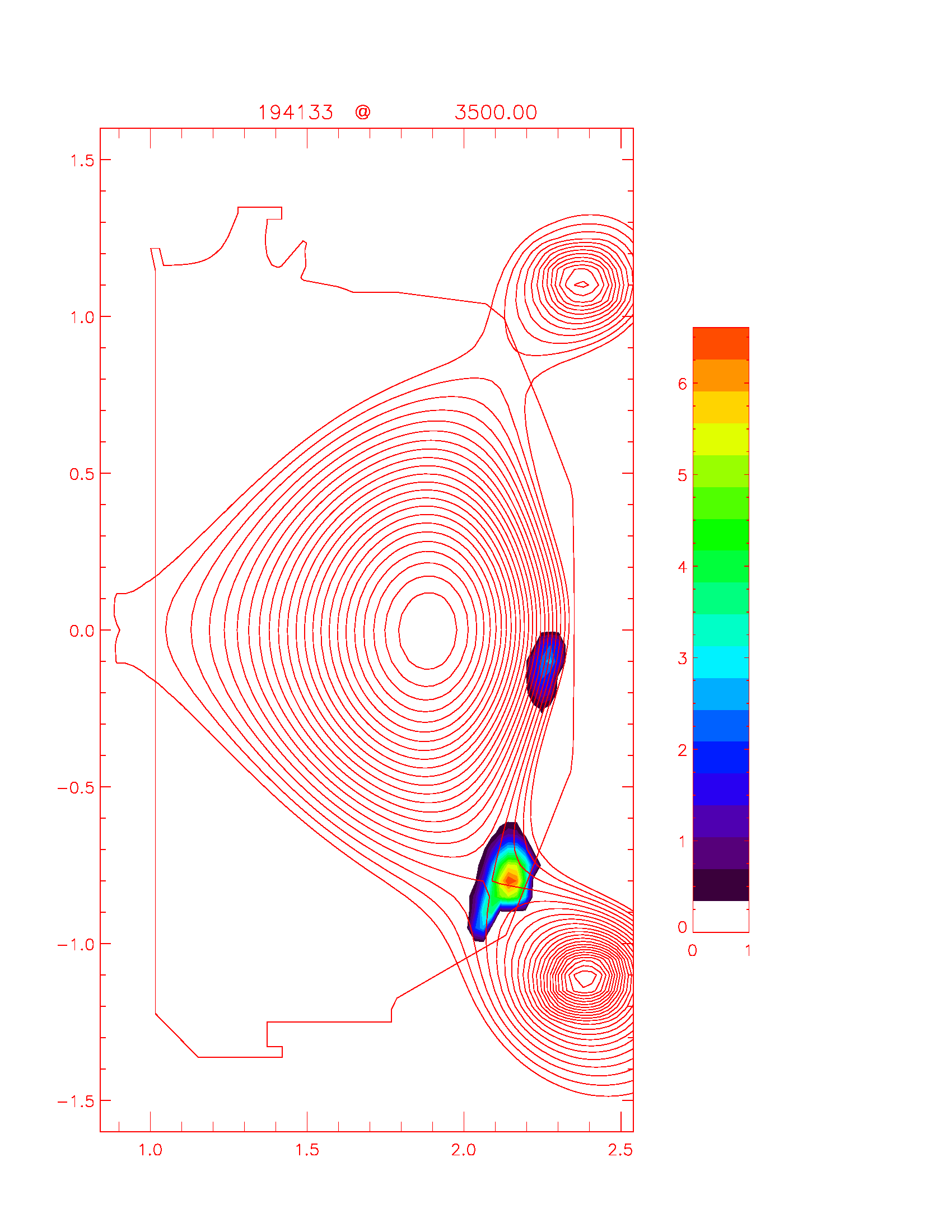}
    \includegraphics[scale=0.35]{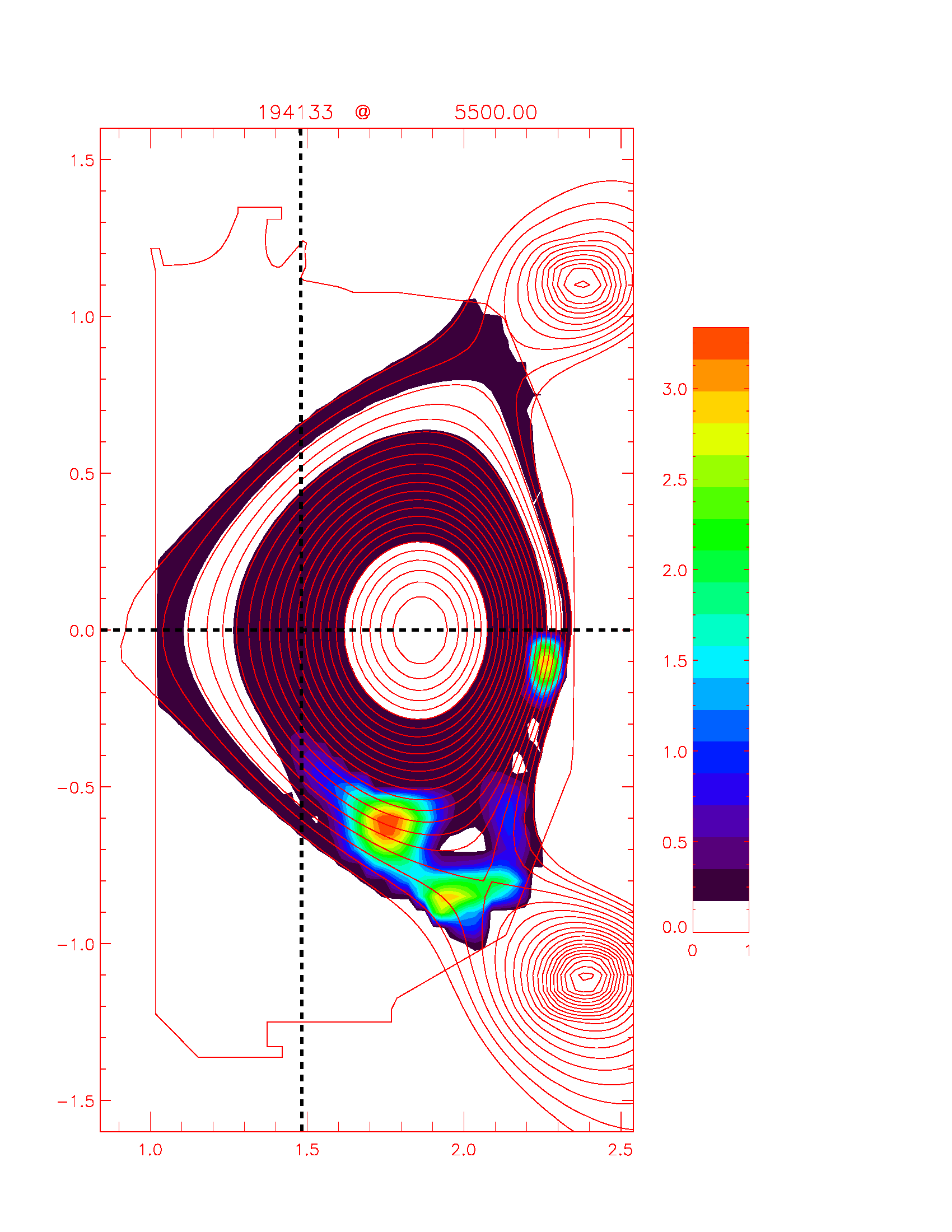}
    \caption{Bolometer inversion for shot 194133 at $t=3500$ and 5500, that is at $f_G=1.15$ and 1.95. On the second figure, we also show the horizontal ``R0'' and vertical ``V1'' interferometer line of sights with black dashed lines.}
    \label{fig2}
\end{figure*}

We show in Fig.\ref{fig1}, density ramps up to the density limit at three different power levels for the $I_p=0.6$MA, magnetic field $B_0=+2$T scenario. The ohmic case reaches a Greenwald fraction $f_G\simeq 1.2$, at $P_{NBI}=3.2$MW $f_G\simeq 1.7$, while at 11.8MW $f_G\simeq 1.95$ is reached and sustained, with $\beta_N\sim2$, until the power drops to 9.5MW and the shot disrupts at $f_G\simeq 2$. $f_G$ (solid lines) shown in Fig.\ref{fig1}a is calculated using the line-averaged density from the horizontal interferometer chord at $Z=0\simeq Z_{axis}$ (``R0'' line of sight) divided by $I_p$[MA]/($\pi\,a$[m]$^2$) in [$10^{20}$m$^{-3}$] units. As noted in \cite{Fevrier2024}, the density in the scrape-off layer (SOL) may not be negligible, especially near the density limits and in particular for vertical lines near the divertor region where the Marfe forms, mainly on the high field side (HFS). The multifaceted asymmetric radiation from the edge (Marfe) is a poloidally localized radiation front which forms near the X-point and expands along the last closed flux surface (LCFS) \cite{Lipschultz_1984}.
Therefore, the line-averaged density calculated from the Thomson scattering profiles mapped along the interferometer chord and divided by the length inside the LCFS is used \cite{Fevrier2024}. We show in dashed lines the Greenwald fraction $f_G^{LCFS}$ obtained with this calculation along the R0 line. First we see that it matches well the interferometer values at low density, meaning that Thomson scattering was well calibrated. Second, at high density, a difference up to $\sim15\%$ can be observed. The maximum Greenwald fraction within the LCFS becomes $f_G^{LCFS}\simeq0.97$ in ohmic, 1.43 at 3.2MW and a value of 1.70 is sustained at 11.8MW. In the latter discharge, 194133, the Marfe starts at about $t=3900$ms and a significant radiated power near the X-point is observed, with large contributions along the LCFS, mainly on the HFS and some on the low field side as well. We show in Fig. \ref{fig2} the bolometer inversion just before the Marfe and detachment start, and once it is well developed where we see the increase of radiated power on the HFS. We also show in Fig. \ref{fig2}b the interferometer horizontal and vertical lines of sight discussed in this paper.

The evolution of the Marfe as well as the edge turbulence properties are studied in detail in the companion paper \cite{Hong2025}. As also discussed in \cite{Scotti2024}, the radiated power inside the LCFS increases when the plasma detaches due to the proximity of the X-point to the DIII-D wall. This radiated power enhances as well the confinement degradation, since it affects the plasma edge inside the LCFS (in 194133, $H_{98y2}$ drops from 1 at $t=3800$ms to 0.7 after 4500ms). We see a clear drop of $\beta_N$ at constant power when this happens (at $t\simeq3600$ in ohmic (194136) and with 3.2MW (194130), and at $t\simeq4000$ at 8MW (194133)). The detachment starts at about $t=3450$ms for 194136 (ohmic), at 3700ms for 194130 (3.2MW) and 3800ms for 194133 (8MW). Dedicated studies with a closed divertor with longer divertor legs will be necessary to determine if the combination of detachment and confinement degradation is inherent to NT plasmas at high density or due to the present experimental set-up. The fact that this combination is as strong at lower power and therefore lower densities tends to indicate that it is not due to high density in NT plasmas per se. We show that NT plasmas can easily sustain $f_G^{LCFS}>>1$ but also can sustain good performance, even if the experimental set-up is unfavourable. This is also demonstrated in shot=194133, where the power is increased from 8MW to 10.5MW at 5000ms, and then to 11.8MW. $\beta_N$ recovers well to $\beta_N\simeq2$ although $f_G$ is above 1.8 ($f_G^{LCFS}>1.6$) for more than 20 total confinement times.

\section{Profiles evolution}
\label{sec:profiles}

\begin{figure*}
    \centering
    \includegraphics[scale=0.5]{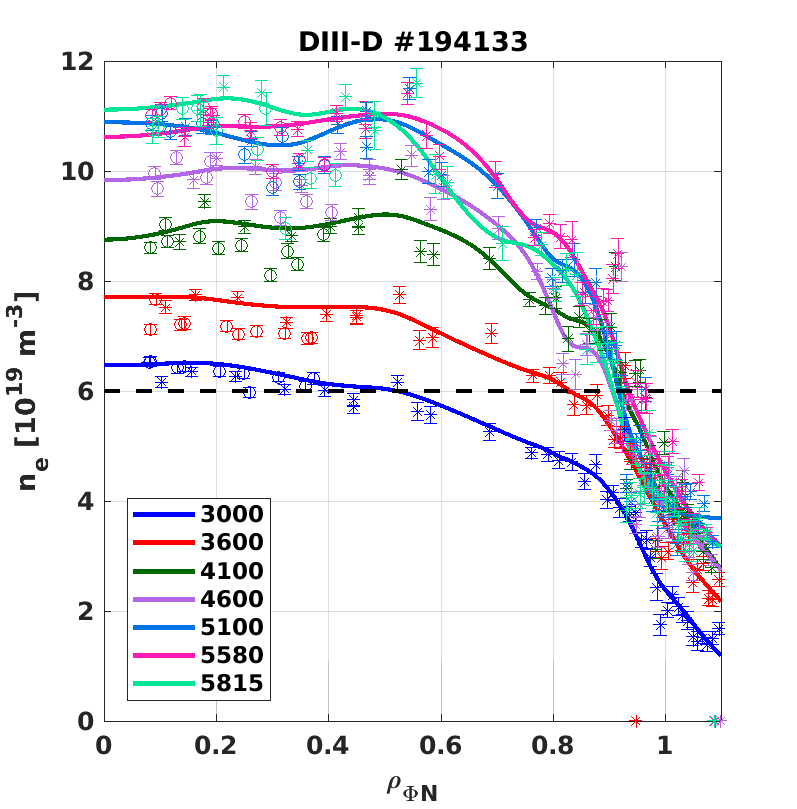}
    \includegraphics[scale=0.5]{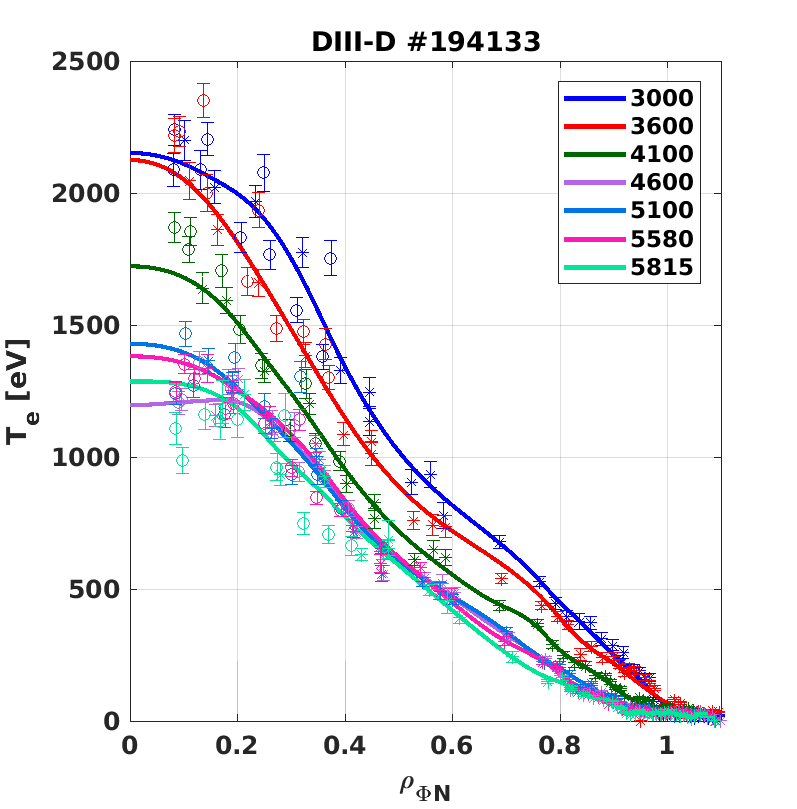}
    \includegraphics[scale=0.5]{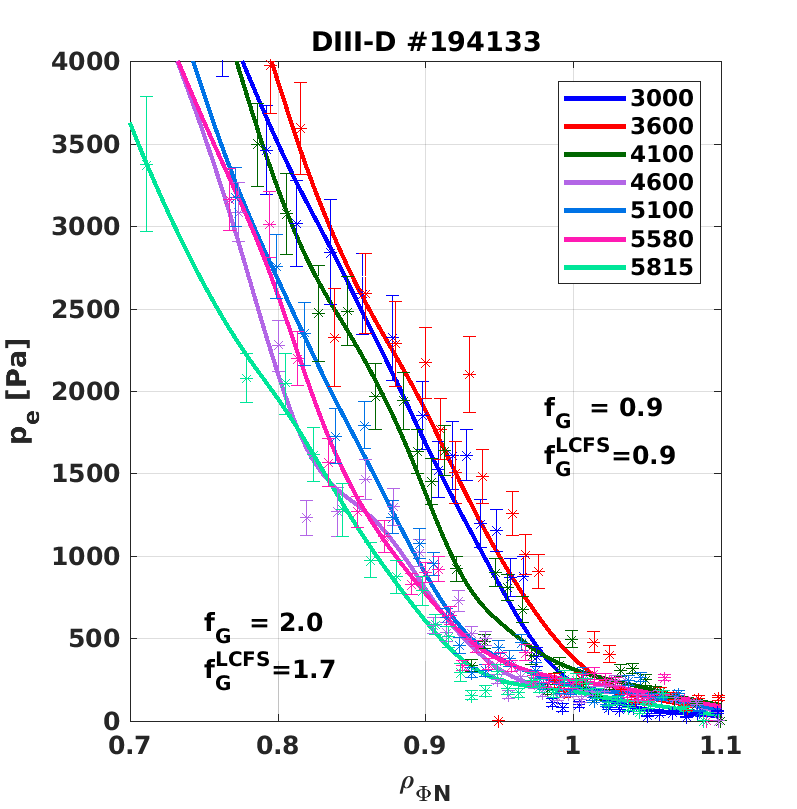}
    \caption{(a) Electron density , (b) temperature and (c) (zoom of) pressure profiles for shot=194133, $I_p=0.6$MA, at $t=3000/f_G=0.92/f_G^{LCFS}=0.91$; 3600/1.15/1.12; 4100/1.5/1.37; 4600/1.7/1.53; 5100/1.85/1.6; 5580/1.95/1.7; 5815/2.0/1.6, respectively. The times are marked in the legends. The fits are based on the core Thomson scattering measurements within $\pm$10ms. The value of the Greenwald density is shown as a horizontal black dashed line in (a).}
    \label{fig3}
\end{figure*}

\begin{figure}[H]
    \centering
    \includegraphics[scale=0.57]{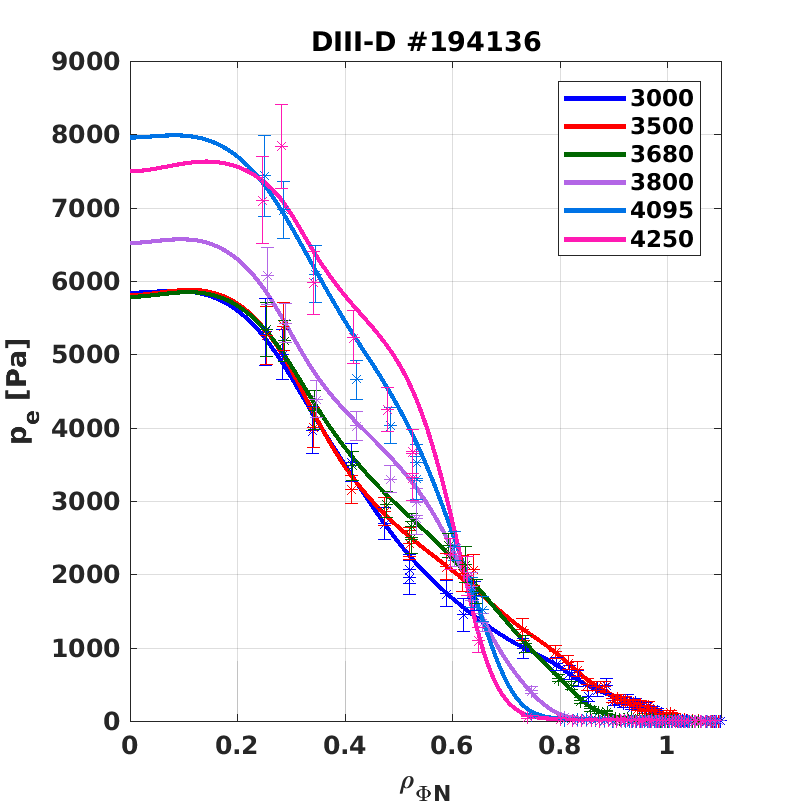}
    \includegraphics[scale=0.57]{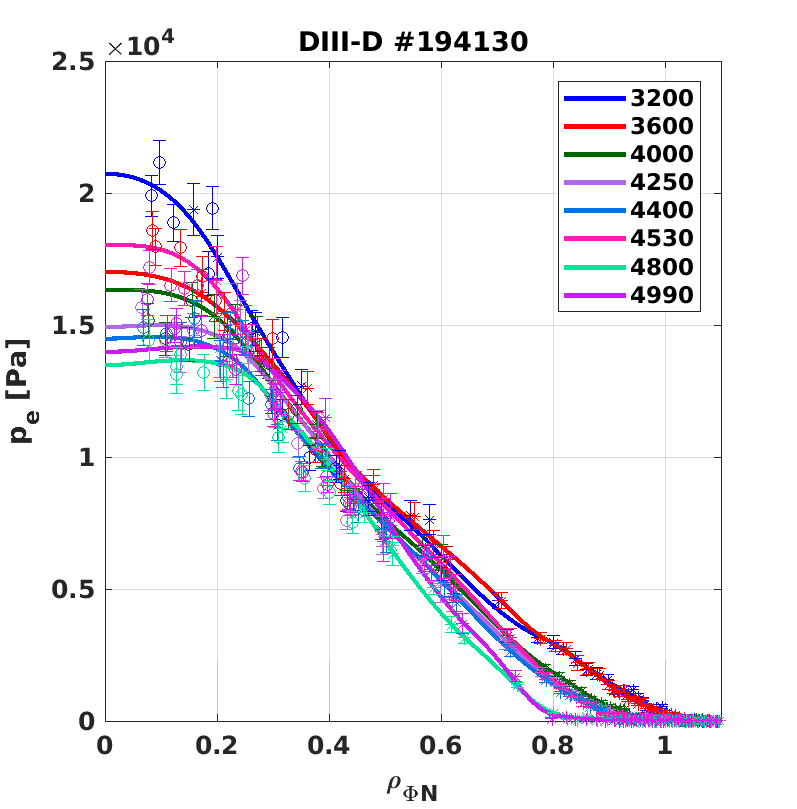}
    \caption{Electron pressure profiles for (a) shot=194136 (ohmic) and (b) 194130 (3.2MW), legends indicate the times in ms. This complements the profiles shown in Fig. \ref{fig3}c for 194133 to cover all cases shown in Fig. \ref{fig1}. Similar pressure profile evolution is observed, albeit at different Greenwald fractions thanks to the auxiliary power.}
    \label{figApp1}
\end{figure}

The natural next step following the results presented in Sec. 2 is to analyze the kinetic profiles and their time evolution up to such large Greenwald fractions. In Fig. \ref{fig3}, we show the profiles of the record shot at $I_p=0.6$MA, 194133, at different times during the relatively slow density ramp, recalling that the power input is slightly increased at 5000ms, as shown in Fig. \ref{fig1}. The value of $n_e=n_{G}$ is shown with a dashed black line in Fig. \ref{fig3}a. We see that from 4100ms onwards, $n_e(\rho_{\Phi N} \leq 0.92) > n_G$, with central values almost twice the Greenwald value. We define $\rho_{\Phi N}$ as the squareroot of the normalized toroidal flux. In terms of squareroot of normalized poloidal flux, it yields $n_e(\rho_{\psi N} \leq 0.97)>n_G$. The value at the separatrix remains slightly below $n_G$; however, we also see that the density profile outside $\rho_{\Phi N}=0.9$ remains essentially constant. The core values continue to increase up to about 5100ms $(f_G=1.85, f_G^{LCFS}=1.6$) and then remain essentially constant($(f_G=1.95, f_G^{LCFS}=1.68$ at $t=5600$ms). This is why the study of the power dependence of the density near the density limit is complex and different between core and edge. This is studied in detail in Ref. \cite{Hong2025}, and discussed with the turbulence measurements. The electron temperature and pressure profiles show a significant flattening for $\rho_{\Phi N}\geq0.9$ from 4600ms onwards. In Fig. \ref{fig3}c, we show a zoom near the edge of the $p_e$ profiles where the evolution for $t>4600$ms shows a slow broadening of the flat edge region. The pressure profiles for the other two shots, 194136 (ohmic) and 193130 (3.2MW), are shown in Fig. \ref{figApp1}. We clearly see that with less auxiliary power, the flat edge pressure profile broadens significantly and at smaller $f_G$. 

The evolution of the kinetic profiles follows the one observed in PT discharges near the density limit across tokamaks, see e.g. Fig. 2 of \cite{Giacomin2022}, and in TCV ohmic NT plasmas (Fig. 21 of \cite{Sauter2014}). The collapse of the edge pressure gradient is associated with a large increase of edge turbulence in \cite{Giacomin2022, Giacomin2022a}, where it is shown to predict the value of the density around 0.9-0.95 at the Marfe onset \cite{Giacomin2022}. The role of turbulence is studied in more detailed in \cite{Hong2025}, and we note that one could expect better resilience in NT to gradient collapse since NT also reduces the SOL turbulence \cite{Lim2024}, hence requiring lower heat fluxes than in PT to sustain finite temperature gradients. However, in the present discharges, we observe such an edge collapse much before the disruption, near the formation of the Marfe and near the detachment time. Usually in PT discharges, with such an edge $T_e$ flattening as shown in Fig. \ref{fig3}b from 5100ms onwards, tearing modes would be triggered and then a disruption soon after, especially at $\beta_N\geq2$. Therefore, dedicated studies should be performed to understand if NT is more resilient than PT to ``edge cooling'', possibly since the edge magnetic shear is fundamentally different between NT and PT \cite{MarinoniReview}. Another aspect to analyze is if the additional heat flux is able to inhibit a local radiation collapse near $q=2$ \cite{Gates, Zanca}. Similar dedicated experiments at constant power in PT discharges can help differentiating these effects. Note that the density ramp near the density limit needs to be relatively slow, since the role of the $q$ profile evolution, due to the modification of the edge temperature, is also important.

\begin{figure*}
    \centering
    \includegraphics[scale=0.57]{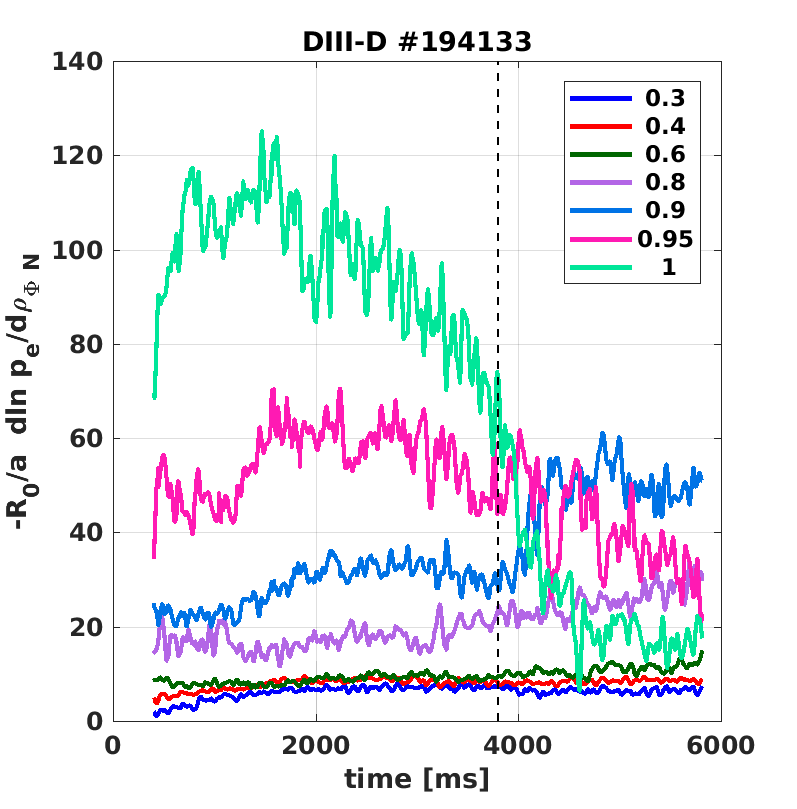}
    \includegraphics[scale=0.57]{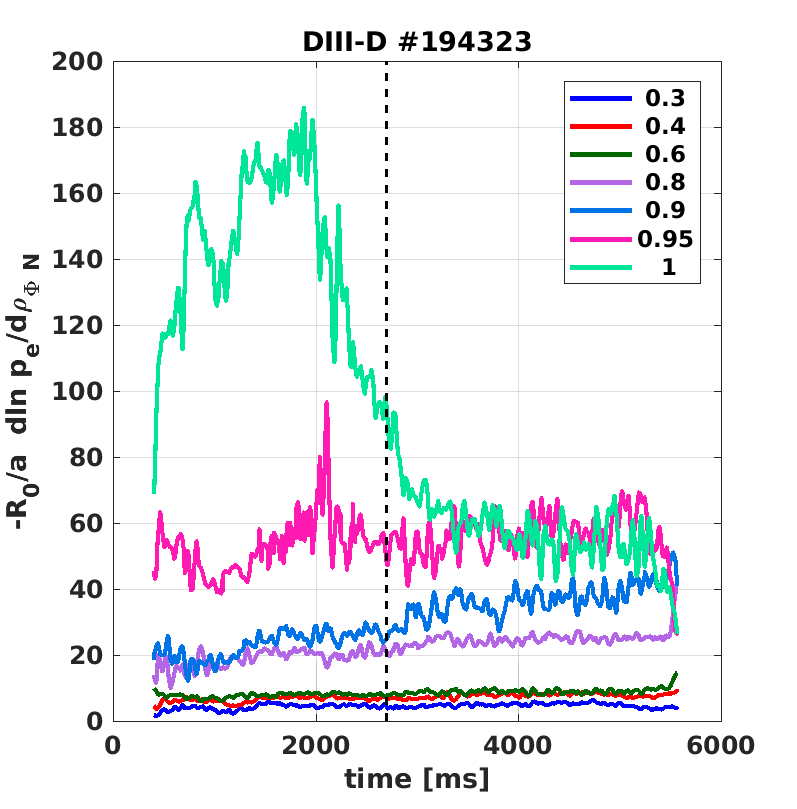}
    \caption{Time evolution of $R_0/L_{pe}$ at different normalized radii (see legend) for (a) shot=194133 (0.6MA, +2T) and (b) shot=194323 (0.8MA, -2T). The vertical dashed lines marked the approximate time of detachment (194133: 3800ms, 194323: 2700ms.}
    \label{fig4}
\end{figure*}

We have shown that the edge electron temperature profile flattens significantly as the density is increased, inducing a broadening of the edge pressure ``collapse'' and a loss of confinement properties. In order to better analyze the evolution of the core plasma transport properties, we show in Fig. \ref{fig4}a the inverse electron pressure scale length of shot 194133 at seven different radii from $\rho_{\Phi N}=0.3$ to 1.0 ($R_0/L_{pe} = - R_0/a\, \cdot \, dln\,p_e/d\rho_{\Phi N}$). First, for $t<4000$ms, we see that the values in the core ( $\rho_{\Phi N}=0.3-0.6$) are relatively similar and similar to values in PT, while the inverse scale length increases significantly towards the edge reaching very high values at  $\rho_{\Phi N}=0.95-1$, as obtained in TCV (\cite{Sauter2014}, Sec. IV). Second, we see the drop of $R_0/L_{pe}$ at  $\rho_{\Phi N}=1$, for $t>4000$, reaching values below the ones at  $\rho_{\Phi N}=0.9$ and 0.95, with a consequent small increase at  $\rho_{\Phi N}=0.9$ explaining how $\beta_N\sim 2$ can be sustained. The case shown in Fig. \ref{fig4}b, with $I_p=0.8$MA, will be discussed in the next Section.

\section{Experimental results at 0.8MA and 1MA}
\label{sec:otherIps}

\begin{figure}[H]
    \centering
    \includegraphics[scale=0.57]{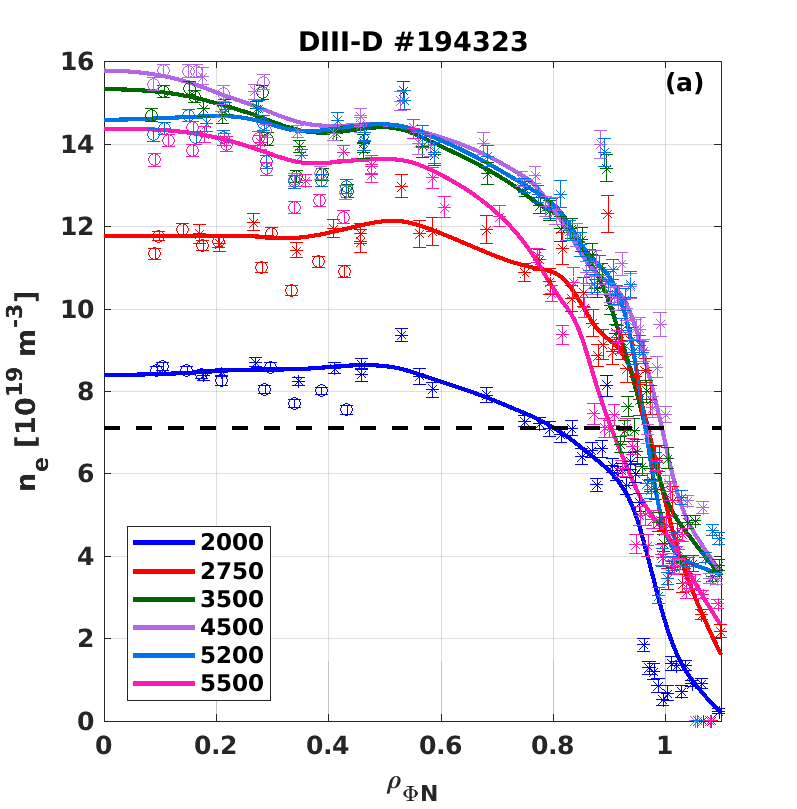}
    \includegraphics[scale=0.57]{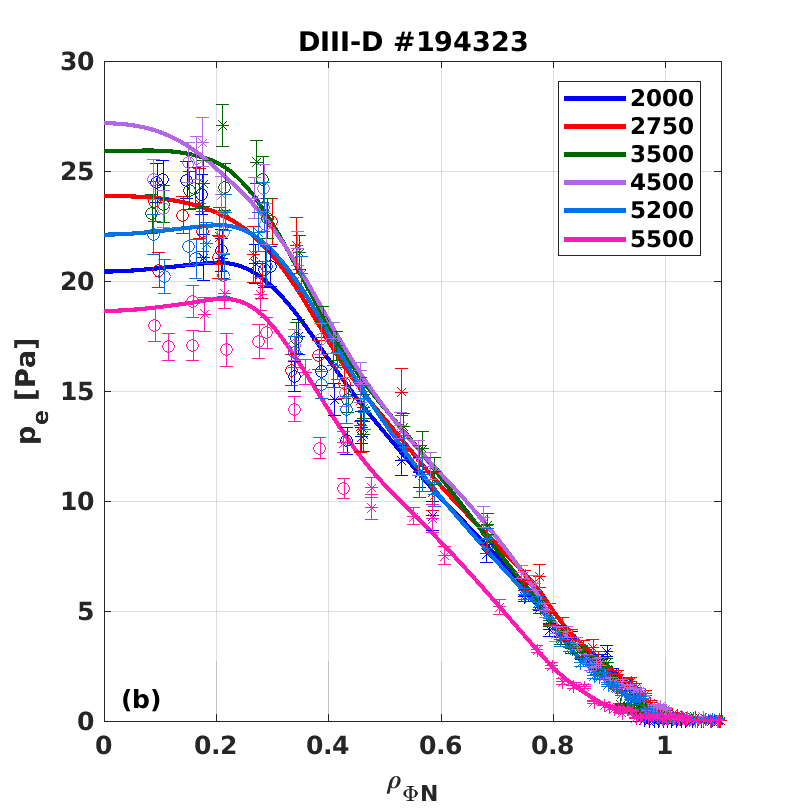}
    \caption{Electron (a) density and (b) pressure profiles for shot=194323. The time traces are shown in Fig. \ref{fig5}b and inverse pressure scale length in Fig. \ref{fig4}b. Note that finite edge pressure gradient is maintained throughout the pulse, up to 5200ms, and we see a flattening during the ramp-down (5500ms) and while the power is not sustained either. Similarly stationary high core density, about twice $n_G$ (horizontal black dashed line in (a)), and separatrix density near $n_G$.}
    \label{figApp2}
\end{figure}

\begin{figure*}
    \centering
    \includegraphics[scale=0.75]{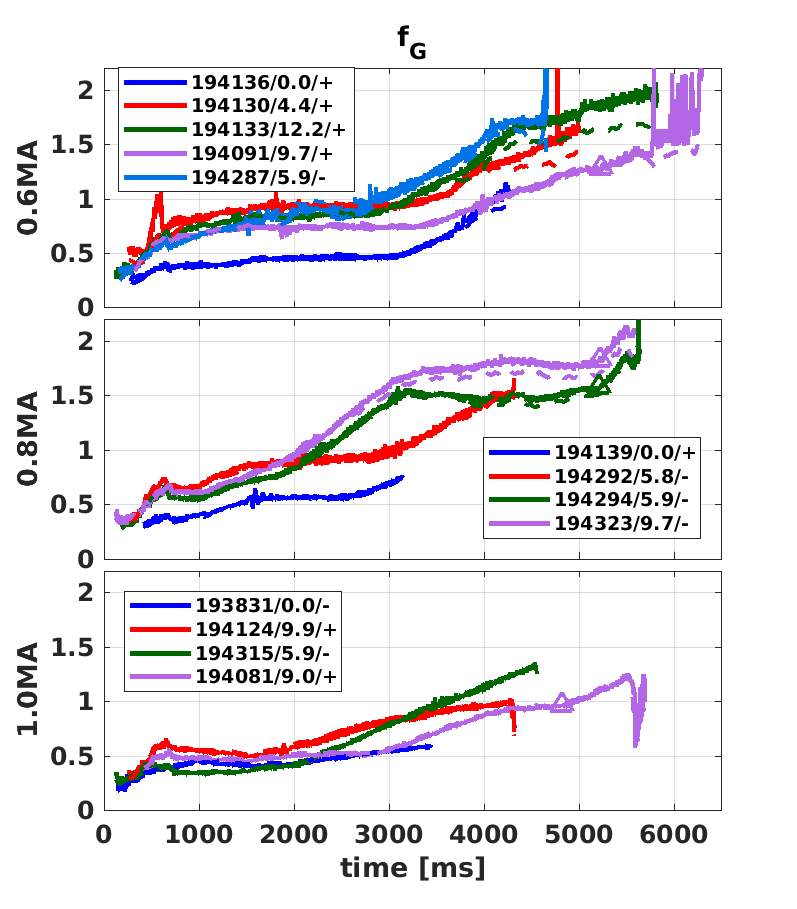}
    \caption{Time evolution of the Greenwald fraction $f_G$ for different plasma currents: (a) 0.6MA, (b) 0.8MA and (c) 1MA. Discharges with $\pm$2T are shown. Negative $B_0$ cases are marked with a minus sign in the legends, and positive with ``+''. The 2nd number in the legends refers to the maximum NBI power. Hence ``194136/0.0/+'' is the ohmic case 194136, 0 NBI, with positive magnetic field. The open triangles mark the start of the current ramp-down phase, when it is reached.}
    \label{fig5}
\end{figure*}

We show in Fig. \ref{fig5} examples of the Greenwald fractions in NT discharges at various plasma currents: 0.6MA ($q_{95}\simeq 4.1$), 0.8MA ($q_{95}\simeq 3.2$), 1.0MA ($q_{95}\simeq 2.5$). For the 0.6MA case, we added two discharges as compared to Fig. \ref{fig1}, one with opposite magnetic field (194287) and one including an $I_p$ ramp-down at high $f_G$ (194091, $f_G\geq 1.3$). The start of the plasma current ramp-down is marked by an open triangle. The Greenwald fraction increases as the current is decreased faster than the density, and the plasma lands safely down to $I_p=0.24$MA at $f_G=1.8$, $q_{95}=8$. In \cite{Scotti2024}, it has been shown that, similarly to PT plasmas, drifts in the SOL play a role in the detachment properties of the NT plasmas, with $B_0<0$ (Fwd $B_0$) requiring a higher density to detach. In terms of the density limit, we did not observe a significant difference, but we did not perform a dedicated experimental study. We show, in Fig. \ref{fig5}, cases with Rev ('+' sign in the legends) and Fwd $B_0$ (minus sign). The cases studied before were all in Rev $B_0$, including 194133 reaching $f_G=1.95/f_G^{LCFS}=1.7$. We see that at each plasma current, high Greenwald fractions are sustained for both Fwd and Rev $B_0$ cases. We show in particular shot 194323 (Fig. \ref{fig5}b), $I_p=0.8$MA and $B_0=-2T$, which stays at $f_G\simeq1.8/f_G^{LCFS}\simeq1.7$ up to the start of the $I_p$ ramp-down ($t=5200$ms) and reaching $f_G\simeq2.1/f_G^{LCFS}\simeq2.0$ during the termination phase at $I_p=0.58$MA at $t=5500$ms, while the NBI power is reduced as well. The absence of a ``density limit'' phenomena is confirmed by the time evolution of the inverse electron pressure scale lengths, Fig. \ref{fig4}b. After the detachment and related decrease of the edge $R_0/L_{pe}$ (at $t\sim2500$ms), the performance can be maintained stationary at $\beta_N\simeq1.8$ with 6-6.5MW from $t=3000$ms to 5500ms, but also with $H_{98y2}\simeq0.7$. We also see that there is a difference between $f_G$ and $f_G^{LCFS}$ starting from 2500ms, which remains at about 5\%. The inverse scale lengths remain stationary with $R_0/L_{pe}(\rho_{\Phi N}=0.95-1)>R_0/L_{pe}(\rho_{\Phi N}=0.9)$ up to the ramp-down phase. The electron density and pressure profiles for shot 194323 are shown in Fig. \ref{figApp2}, showing no edge flattening up to the end of the flat top (t=5200ms). During the whole flat top, very high density is sustained as well as finite edge pressure gradient.

\begin{figure}[H]
    \centering
    \includegraphics[scale=0.75]{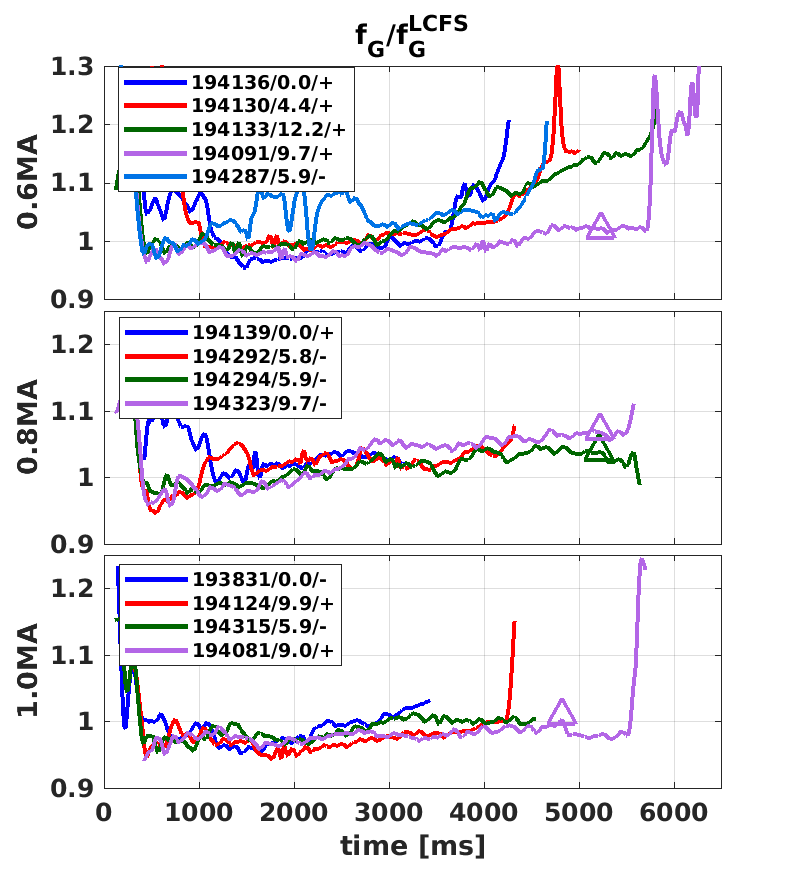}
    \caption{Ratio $f_G/f_G^{LCFS}$ along the ``R0'' line of sight for the same shots as in Fig. \ref{fig5}. The sharp increase is due to fringe jumps or ill-defined interferometer values near the disruption. The open triangles mark the start of the current ramp-down phase, when it is reached.}
    \label{figApp3}
\end{figure}

Consistently with the results presented earlier, the maximum Greenwald fraction obtained in ohmic plasmas is always much lower than with additional auxiliary heating, including at 1MA. Nevertheless, the 1MA discharges are not terminated because of the density limit, since no Marfe formation nor edge pressure collapse are observed. Nonetheless, $f_G$ up to 1.25 have been obtained both in $I_p$ flat top (194315) and during ramp-down (194081), Fig. \ref{fig4}c. We also see that $f_G=f_G^{LCFS}$ throughout at 1MA (dashed lines not visible in Fig. \ref{fig5}c), confirming the absence of high density in the SOL near the $Z=0$ position ($Z=Z_{axis}$). Additional dedicated experiments at 0.8MA are also required. Indeed the density limit is so high that it was not reached during these experiments, due to lack of experimental time after a few trials. We can only state that it is $\gtrsim 2$. The ``closeness'' to the density limit can also be gauged by looking at the difference between $f_G$ and $f_G^{LCFS}$, which relates to a large increase of $n_e$ in the SOL up to $Z=0$ (leading to the change in turbulence properties in the SOL \cite{Giacomin2022, Giacomin2022a}). With $I_p=0.6$MA, we start to see more than 5\% difference for $f_G>1.3$ and for $I_p=0.8$MA when $f_G>1.5$. While for $I_p=1.0$MA and $f_G$ up to 1.2, the ratio remains at 1 (see Fig. \ref{figApp3}). Note that on TCV, using the density profile observer \cite{Pastore2023}, one follows in real-time this ratio as an early indicator of the Marfe formation and the proximity to a density limit. This is particularly true for vertical interferometer chords near the X-point. It was already observed in TCV, that a significant plasma density can be present in SOL on the high field side of the X-point \cite{Coda_2022}. Using the ``V1'' vertical line of sight, located at $R=1.48$m at DIII-D, we observe ratios up to $f_G/f_G^{LCFS}\sim1.6$ (for shot 194133). It should be emphasized that previous worldwide experimental analyses used the line-averaged density provided by the interferometer diagnostic to compute the Greenwald fractions, to our knowledge, and often with a vertical line of sight near the magnetic axis, hence near the X-point location.

\section{Conclusions}
\label{sec:conclusion}

We have demonstrated that NT discharges can reach very high Greenwald fractions, up to values near 2. This is only true with auxiliary heating, since in ohmic discharges, $f_G<1.2$ was observed at all plasma currents. At $I_p=0.6$MA, $B_0=+2$T, a clear power dependence of the density limit is demonstrated for the first time, with $f_G=1.2$ in ohmic, 1.7 at 3.2MW and 1.95 at 11.8MW. The core density can be up to twice the Greenwald density and keeps increasing, while the value at the separatrix remains essentially constant and slightly below $n_G$, with $n_e> n_G$ for $\rho_{\Phi N}\leq0.92$ ($\rho_{\psi N}\leq0.97$). Since the X-point is very close to the wall in our experimental set-up, detachment and Marfe formation affect the whole SOL, leading to significant density in the SOL even up to $Z=Z_{axis}$. Therefore, the ``R0'' horizontal interferometer line of sight is affected and measures higher values than obtained with considering only the part inside the LCFS. The latter line-averaged density is calculated using the Thomson scattering profile. Thomson scattering measurements should be used for density limit/detachment studies, as proposed in \cite{Fevrier2024}. Using this method, the values mentioned above become $f_G^{LCFS}=0.97$ in ohmic, 1.43 at 3.2MW and 1.7 at 11.8MW. These remain record values, to our knowledge, in gas puff tokamak discharges and have been obtained up to $\beta_N\geq2$ (see also \cite{Paz-Soldan2024}). We did not have time to test with additional pellet injection near the density limit, which are known to allow for a further increase of the core density \cite{Kamada_1991}. From the present results, Greenwald fraction of about 1.5 as a target for FPP (fusion power plant) design seems reasonable for NT plasmas. Of course, testing if this requires a certain strong shaping will be the subject of future studies. It clearly provides an attractive FPP solution for low q, high density, high confinement NT-plasmas, complementary to the recent high q, high density, high confinement PT-H modes, results obtained in DIII-D \cite{Ding2024}.

Discharges with Fwd and Rev $B_0$ orientations have been analyzed and shown to reach very high Greenwald fractions. Our LCFS shape has sufficiently NT values to inhibit access to H-modes in both Fwd and Rev magnetic field directions \cite{NelsonPRL}. At 0.8MA, -2T, $q_{95}=3.2$, a stable detached stationary discharge is sustained with 6-6.5MW NBI at $\beta_N\geq1.8$ and $f_G=1.8$, $f_G^{LCFS}=1.7$. This is very promising for NT fusion reactors, since high density is necessary. In our case, as described in \cite{Scotti2024}, the confinement properties drop to about $H_{98y2}=0.7$ after the plasma detaches (from values of unity just before detachment). Experiments with a closed divertor ``distant'' from the X-point are required to test if the confinement degradation is due to our specific geometry or inherent to high density NT plasmas.

\ack
This work has been carried out in part within the framework of the EUROfusion Consortium, via the Euratom Research and Training Programme (Grant Agreement No 101052200 — EUROfusion) and funded by the Swiss State Secretariat for Education, Research and Innovation (SERI). Views and opinions expressed are however those of the author(s) only and do not necessarily reflect those of the European Union, the European Commission, or SERI. Neither the European Union nor the European Commission nor SERI can be held responsible for them.
This work was supported in part by the US Department of Energy under the following awards DE-FC02-04ER54698, DE-AC52-07NA27344, DE-AC05-00OR22725, DE-SC0022270, DE-SC0019352, DE-FG02-08ER54999. The author (R.H.) performed the work supported by the U.S. Department of Energy, Office of Science, Office of Fusion Energy Sciences, under Award DE-SC0019352.

\section*{Data availability statement}
This is the data availability statement required by policies of the DIII-D National Fusion Facility. The data that support the findings of this study are available upon reasonable request
from the authors.

\section*{Disclaimer}
This report was prepared as an account of work sponsored by an agency of the United States Government.  Neither the United States Government nor any agency thereof, nor any of their employees, makes any warranty, express or implied, or assumes any legal liability or responsibility for the accuracy, completeness, or usefulness of any information, apparatus, product, or process disclosed, or represents that its use would not infringe privately owned rights.  Reference herein to any specific commercial product, process, or service by trade name, trademark, manufacturer, or otherwise, does not necessarily constitute or imply its endorsement, recommendation, or favoring by the United States Government or any agency thereof.  The views and opinions of authors expressed herein do not necessarily state or reflect those of the United States Government or any agency thereof.
\\

\printbibliography
%\bibliography{myreferences}

\end{document}